\documentclass[twocolumn,english,reprint, longbibliography, superscriptaddress, breaklinks=true, showkeys, showpacs=false, nofootinbib]{revtex4}
\usepackage[T1]{fontenc}
\usepackage[latin9]{inputenc}
\usepackage{color}
\usepackage{babel}
\usepackage{amsmath}
\usepackage{amssymb}
\usepackage{subfigure}
\usepackage{graphicx}
\usepackage{physics} 
\usepackage{hyperref}
\hypersetup{
    colorlinks=true,
    linkcolor=blue,
    filecolor=blue,      
    urlcolor=blue,
    }
\makeatletter
\@ifundefined{textcolor}{}
{%
 \definecolor{BLACK}{gray}{0}
 \definecolor{WHITE}{gray}{1}
 \definecolor{RED}{rgb}{1,0,0}
 \definecolor{GREEN}{rgb}{0,1,0}
 \definecolor{BLUE}{rgb}{0,0,1}
 \definecolor{CYAN}{cmyk}{1,0,0,0}
 \definecolor{MAGENTA}{cmyk}{0,1,0,0}
 \definecolor{YELLOW}{cmyk}{0,0,1,0}
}
\pdfoutput=1
\hypersetup{colorlinks=true,citecolor=blue,linkcolor=cyan,urlcolor=blue,filecolor= green, breaklinks=true}
\usepackage{url}
\usepackage{breakurl}
\makeatother
\usepackage{qcircuit}

\begin{document}

%\title{Comment on ``Implementation of a general single-qubit positive operator-valued measure on a circuit-based quantum computer''}
\title{Simulation of positive operator-valued measures and quantum instruments \\ via quantum state preparation algorithms}

\author{Douglas F. Pinto}
\email{douglasfpinto@gmail.com}
\address{Physics Department, Center for Natural and Exact Sciences, Federal University of Santa Maria, Roraima Avenue 1000, Santa Maria, Rio Grande do Sul, 97105-900, Brazil}

\author{Marcelo S. Zanetti}
\email{marcelo.zanetti@ufsm.br}
\address{Department of Electronics and Computing, Technology Center, Federal University of Santa Maria, Roraima Avenue 1000, Santa Maria, Rio Grande do Sul, 97105-900, Brazil}

\author{Marcos L. W. Basso}
\email{marcoslwbasso@hotmail.com}
\address{Center for Natural and Human Sciences, Federal University of the ABC, States Avenue 5001, Santo Andr\'e, S\~ao Paulo, 09210-580, Brazil}

\author{Jonas Maziero}
\email{jonas.maziero@ufsm.br}
\address{Physics Department, Center for Natural and Exact Sciences, Federal University of Santa Maria, Roraima Avenue 1000, Santa Maria, Rio Grande do Sul, 97105-900, Brazil}

\begin{abstract}
In Ref. [Phys. Rev. A 100, 062317 (2019)], the authors reported an algorithm to implement, in a circuit-based quantum computer, a general quantum measurement (GQM) of a two-level quantum system, a qubit. Even though their algorithm seems right, 
its application involves the solution of an intricate non-linear system of equations in order to obtain the angles determining the quantum circuit to be implemented for the simulation. In this article, we identify and discuss a simple way to circumvent this issue and implement GQMs on any $d$-level quantum system through quantum state preparation algorithms. Using some examples for one qubit, one qutrit and two qubits, we illustrate the easy of application of our protocol. 
Besides, we show how one can utilize our protocol for simulating quantum instruments, for which we also give an example. All our examples are demonstrated using IBM's quantum processors. 
\end{abstract}


\maketitle

% POVM RELEVANCE AND FORMALISM
% POVM IMPLEMENTATION ON CBQC (REMOVIDO)
% THE PAPER WE COMMENT ON
% THE COMMENT STRATEGY AND METHOD 
% EXAMPLES COMPARING OUR STRATEGY AND THEIR STRATEGY IN TERMS OF "TOMOGRAPHY", AUXILIARY Q-BITS AND CIRCUIT DEPTH (COMPLEXITY)

% POVM RELEVANCE AND FORMALISM

One of the basic postulates of quantum mechanics says that the measurement of an observable, represented by an hermitian operator $A=\sum_{j=1}^{d_{A}}a_{j}\Pi_{j}^{A}$, is described by projection operators $\Pi_{j}^{A}$, i.e., $\Pi_{j}^{A}\Pi_{k}^{A}=\delta_{jk}\Pi_{j}^{A}$ and $\sum_{j=1}^{d_{A}}\Pi_{j}^{A}=\mathbb{I}_{A}$, with $d_{A}$ being the dimension of the system Hilbert space $\mathcal{H}_{A}$ and $\mathbb{I}_{A}$ is the identity operator on $\mathcal{H}_{A}$. 
For a quantum system $A$ prepared in the state $\rho_{A}$, the measurement outcome corresponding to $\Pi_{j}^{A}$ is obtained with probability $Pr(\Pi_{j}^{A}|\rho_{A})=\Tr(\Pi_{j}^{A}\rho_{A}\Pi_{j}^{A})$ and the post-measurement state is $\Pi_{j}^{A}\rho_{A}\Pi_{j}^{A}/\Tr(\Pi_{j}^{A}\rho_{A}\Pi_{j}^{A})$.
Eventually, researchers realized that more general quantum measurements (GQMs) can be defined. 
These measurements, also named positive operator-value measurements (POVMs), are described by a set of measurement operators $\{M_{j}\}$ in $\mathcal{H}_{A}$ that satisfy the completeness relation $\sum_{j}M_{j}^{\dagger}M_{j}=\mathbb{I}_{A}$. 
In this general setting, for a system prepared in the state $\rho_{A}$, the probability of obtaining the measurement result corresponding to $M_{j}$ is $Pr(M_{j}|\rho_{A})=\Tr(M_{j}\rho_{A}M_{j}^{\dagger})$ and %, if $M_{j}$ is obtained,
the post-measurement state is $M_{j}\rho_{A}M_{j}^{\dagger}/\Tr(M_{j}\rho_{A}M_{j}^{\dagger})$. The fact that $Pr(M_{j}|\rho_{A})=\Tr(E_{j}\rho_{A})\ge0$  with $E_{j}=M_{j}^{\dagger}M_{j}$ 
being positive-semidefinite operators motivates the name POVM. The completeness restriction ensures that $\sum_{j}Pr(M_{j}|\rho_{A})=1$ \cite{Nielsen,Wilde}.

POVMs provide advantages in several applications in Quantum Information Science (QIS), as for example in quantum state estimation \cite{QEE}, shadow quantum state tomography \cite{nguyen}, discrimination of quantum states \cite{Bergou}, randomness certification \cite{rand}, acquisition of information from a quantum source \cite{info}, quantum key distribution \cite{qkd}, Bell inequalities \cite{bell}, and device independent quantum information protocols \cite{diqi}. 
So, experimentally implementing POVMs is of fundamental importance for QIS and considerable work has been done in this direction recently, as for example in Refs. \cite{Ahnert,ziman,liu,bian,zhao,oszmaniec,li,oszmaniec2,hou,Yordanov}.

% POVM IMPLEMENTATION ON CBQC
%There have been several attemps of POVM implementation in the literature, including $\dots$ {\color{red}{[citation]}}, $\dots$ {\color{red}{[citation]}}, $\dots$ {\color{red}{[citation]}}, $\dots$ and circuit-based quantum computers (CBQC). Besides allowing for the development of the GQM formalism, a CBQC implementation provides an affordable validation approach through experiments on simulated and real hardware {\color{red}{[citation]}}. 

% THE PAPER WE COMMENT ON
Of particular interest to us here is Ref. \cite{Yordanov}, where the authors proposed a deterministic protocol to implement single-qubit POVMs on quantum computers.   
Even though their protocol seems correct, 
%we identify some basic errors on the examples presented by the authors. 
%In addition to that, 
we realize that for applying it  one first has to solve complicated non-linear systems of equations for obtaining the angles determining the quantum circuit to be used in the simulation. 
Then, motivated by their work, here we identify and discuss a simple way to implement POVMs on any $d_{A}$-level quantum system through quantum state preparation (QSP) algorithms \cite{kitaev,shende,plesch,arrazola,araujo,he,zhang,veras}. 
Using some examples for $d_{A}=2$, $d_{A}=3$, and $d_A=4$ we illustrate the simplicity and convenience for application of this new method. % and we also show that it is favorable in terms of the involved auxiliary qubits and quantum circuit depth.
In addition to that, we apply our protocol for the simulation of quantum instruments \cite{Wilde}.

A POVM with elements $\{M_{j}\}$ can be implemented coherently through an isometric transformation \cite{Wilde},
\begin{equation}
V_{AB}|k\rangle_{A}\otimes|0\rangle_{B} := \sum_{j}(M_{j}|k\rangle_{A})\otimes|j-1\rangle_{B},
\label{eq:iso}
\end{equation}
followed by a selective projective measurement in the basis $\{|j\rangle_{B}\}$ of the auxiliary system $B$, plus discarding of the system $B$. Above $|j\rangle_S$ is the computational basis for the system $S=A,B$. This procedure produces the same statistics and post-measurement states of the system $A$ as does the POVM $\{M_{j}\}$, that is to say, $Pr(|j\rangle_{B}|\tilde{\rho}_{AB})=Pr(M_{j}|\rho_{A})$ with $\tilde{\rho}_{AB}=V_{AB}(\rho_{A}\otimes|0\rangle_{B}\langle0|)V_{AB}^{\dagger}$ and $Tr_B \big(\mathbb{I}_{A}\otimes|j\rangle_{B}\langle j|\big)\tilde{\rho}_{AB}\big(\mathbb{I}_{A}\otimes|j\rangle_{B}\langle j|\big)/\Tr\big(\big(\mathbb{I}_{A}\otimes|j\rangle_{B}\langle j|\big)\tilde{\rho}_{AB}\big(\mathbb{I}_{A}\otimes|j\rangle_{B}\langle j|\big)\big) = M_{j}\rho_{A}M_{j}^{\dagger}/\Tr\big(M_{j}\rho_{A}M_{j}^{\dagger}\big)$, with $\rho_{A}$ being the pre-measurement state of system $A$. So, attempts to implement POVMs experimentally usually start from Eq. (\ref{eq:iso}).

%Here, for didactic purposes, we discuss the protocol a 
%The authors start giving a quantum circuit for performing a 2-element POVM on one-qubit system. 
%The generalization to $n$-element POVM can be found in \cite{Yordanov}. 
%Given the initial state  $\ket{\Psi}_{AB} = \ket{\psi_0}_A \otimes \ket{0}_B = (a \ket{0} + b \ket{1}) \otimes \ket{0}$ for the system $A$ plus the ancilla $B$, the authors apply an unitary operation $U_A$ on the system $A$ such that $\ket{\Psi}_{AB} \to U_A(\ket{\psi_0}_A \otimes \ket{0}_B) = (a' \ket{0} + b' \ket{1}) \otimes \ket{0}$. Afterwards, two controlled $y$-rotations are performed, with the ancilla $B$ as the target and system $A$ as the control qubit. The
%rotations are specified by the angles $\theta_1$ and $\theta_2$ with the states $\ket{0}$ and $\ket{1}$ as activators, respectively. In this stage, the state of system is given by $\ket{\Psi}_{AB} = (D_1 U_A \ket{\psi_0}_A)\otimes \ket{0}_B + (D_2 U_A \ket{\psi_0}_A)\otimes \ket{1}_B$, where $D_1 := \cos \theta_1 \ketbra{0} + \cos \theta_2 \ketbra{1}$ and $D_1 = \sin \theta_1 \ketbra{0} + \sin \theta_2 \ketbra{1}$. Finally, two arbitrary controlled unitary operations $V_1$ and $V_2$ are applied such that
%\begin{align}
    %\ket{\Psi_f}_{AB} = (V_1 D_1 U_A \ket{\psi_0}_A) \ket{0}_B + (V_2 D_2 U_A \ket{\psi_0}_A) \ket{1}_B,
%\end{align}
%which defines the elements of the POVM as $M_j = V_j D_j U_A$ for $j = 1,2.$ It's worth mentioning that, by  the singular value decomposition theorem, the operators $M_1$ and $M_2$ are in the most general form. 
%From this basic building block, 

\begin{figure*}[t!]
\centering
\includegraphics[width=0.95\textwidth]{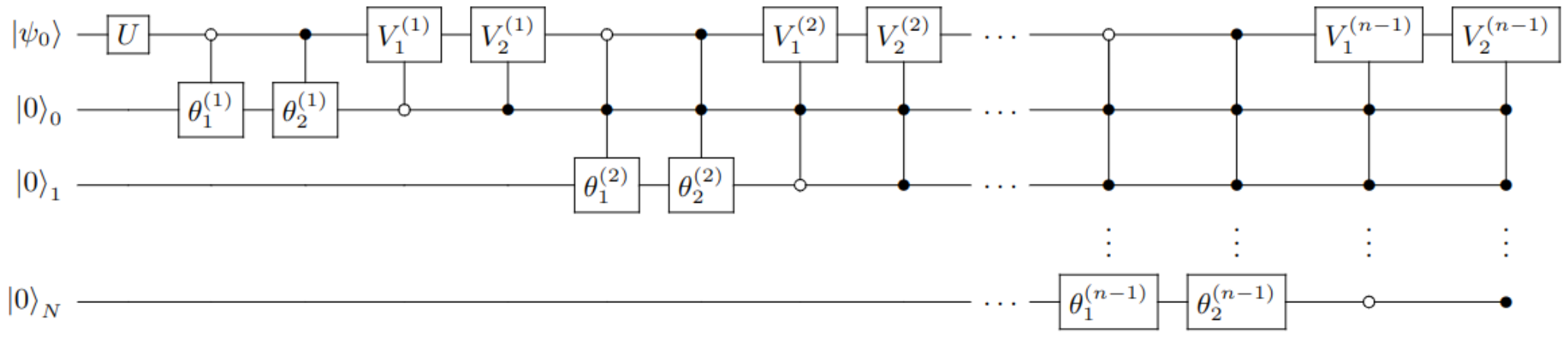}
%\begin{center}
%\mbox{
%\Qcircuit @C=1em @R=.7em {
%\lstick{\ket{\psi_0}} & \gate{U} &\ctrlo{1} &\ctrl{1} & \gate{V_1^{(1)}} & \gate{V_2^{(1)}} &\ctrlo{1} &\ctrl{1} & \gate{V_1^{(2)}}&\gate{V_2^{(2)}}&\qw&\dots& &\ctrlo{1} &\ctrl{1} &\gate{V_1^{(n-1)}} & \gate{V_2^{(n-1)}} & \rstick{ }\\
%\lstick{\ket{0}_0} & \qw &\gate{\theta_1^{(1)}} &\gate{\theta_2^{(1)}} & \ctrlo{-1} & \ctrl{-1} &\ctrl{1} &\ctrl{1} & \ctrl{-1}&\ctrl{-1}&\qw&\dots& &\ctrl{-1} &\ctrl{-1} &\ctrl{-1} & \ctrl{-1} & \rstick{ }\\
% \lstick{\ket{0}_1}& \qw &\qw &\qw & \qw & \qw &\gate{\theta_1^{(2)}}  &\gate{\theta_2^{(2)}}  & \ctrlo{-1}&\ctrl{-1}&\qw&\dots& &\ctrl{-1} &\ctrl{-1} &\ctrl{-1} & \ctrl{-1} & \rstick{ }\\
%  &  & & &  &  &  &  & & & & & &\vdots&\vdots  &\vdots & \vdots & \rstick{ }\\
%   &  & & &  &  &  &  & & & & & &&  & &  & \\
%\lstick{\ket{0}_{N}} & \qw &\qw &\qw & \qw & \qw &\qw  &\qw  & \qw&\qw&\qw&\dots& &\gate{\theta_1^{(n-1)}} &\gate{\theta_2^{(n-1)}}  &\ctrlo{1} & \ctrl{1} & \rstick{ }\\
% & \ctrl{2} & \targ & \gate{U} & \qw& \dots \\
% & \qw & \ctrl{-1} & \qw & \qw & \dots\\
% & \targ & \ctrl{-1} & \ctrl{-2} & \qw & \dots\\
% & \qw & \ctrl{-1} & \qw & \qw& \dots
%}
%}
%\end{center}
\caption{Adapted from the quantum circuit reported in Ref. \cite{Yordanov} to simulate general one-qubit POVMs. $U$ and $V_{j}^{(k)}$ are general one-qubit gates and $\theta_{j}^{(k)}$ represents the $R_{y}(\theta_{j}^{(k)})$ gate. We also used $N=\log_{2}n$.}
\label{fig_yordanov}
\end{figure*}

The protocol given in Ref. \cite{Yordanov} follows this path.
The authors said that the quantum circuit shown in Fig. \ref{fig_yordanov}
prepares the state
\begin{equation}
|\Psi\rangle = \sum_{j=1}^{n-1}(M_{j}|\psi_{0}\rangle)\otimes|o_{1}^{(j)}\rangle + (M_{n}|\psi_{0}\rangle)\otimes|o_{2}^{(n-1)}\rangle,
\label{eq:PsiAB}
\end{equation}
with $|\psi_{0}\rangle$ being the system $A$ pre-measurement state and $\{|o_{1}^{(j)}\rangle\},|o_{2}^{(n-1)}\rangle\}$ are orthonormal states of the auxiliary system $B$, thus implementing a one-qubit POVM with an arbitrary number $n$ of elements:
\begin{equation}
M_{j} = 
\begin{cases} 
V_{1}^{(1)}D_{1}^{(1)}U \text{, para } j=1, \\
V_{1}^{(j)}D_{1}^{(j)}\Pi_{k=1}^{j-1}V_{2}^{(k)}D_{2}^{(k)}U \text{, para } 1<j<n, \\
\Pi_{k=1}^{n-1}V_{2}^{(k)}D_{2}^{(k)}U \text{, para } j = n.
\end{cases}
\label{eq:Mjs}
\end{equation}
In theses equations, $U$ and $V_{j}^{(k)}$ are general one-qubit unitaries and $D_{1}^{(k)}=\cos\theta_{1}^{(k)}|0\rangle\langle 0|+\cos\theta_{2}^{(k)}|1\rangle\langle 1|$  and $D_{2}^{(k)}=\sin\theta_{1}^{(k)}|0\rangle\langle 0|+\sin\theta_{2}^{(k)}|1\rangle\langle 1|$ are positive operators if $\theta_{1}^{(k)},\theta_{2}^{(k)}\in[0,\pi/2]$. 

\begin{figure*}[t!]
\centering
\includegraphics[width=0.70\textwidth]{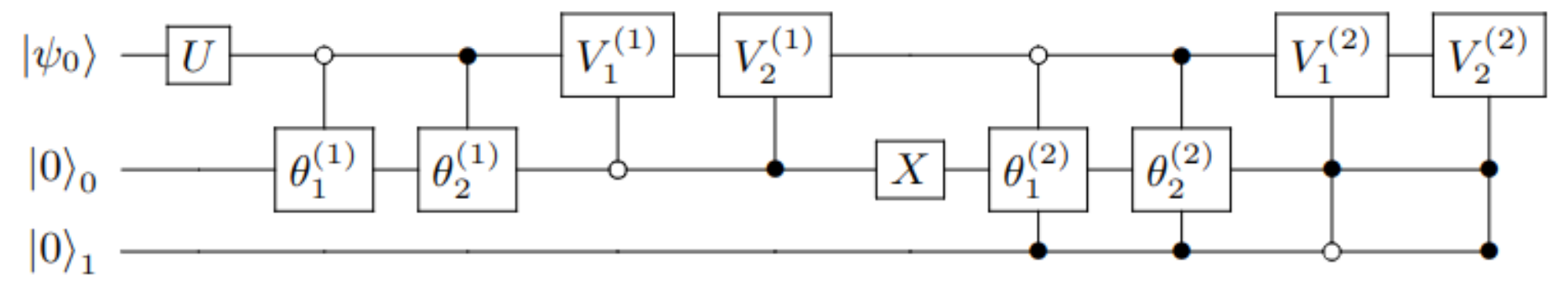}
%\begin{center}
%\mbox{
%\Qcircuit @C=1em @R=.7em {
%\lstick{\ket{\psi_0}} & \gate{U} &\ctrlo{1} &\ctrl{1} & \gate{V_1^{(1)}} & \gate{V_2^{(1)}} & \qw &\ctrlo{1} &\ctrl{1} & \gate{V_1^{(2)}}&\gate{V_2^{(2)}} \\
%\lstick{\ket{0}_0} & \qw &\gate{\theta_1^{(1)}} &\gate{\theta_2^{(1)}} & \ctrlo{-1} & \ctrl{-1} &\gate{X} & \gate{\theta_1^{(2)}} & \gate{\theta_2^{(2)}} & \ctrl{-1}&\ctrl{-1} \\
%\lstick{\ket{0}_1}& \qw &\qw &\qw & \qw & \qw & \qw & \ctrl{-1}  & \ctrl{-1}  & \ctrlo{-1}&\ctrl{-1}
%}
%}
%\end{center}
\caption{Adaptation of the quantum circuit of Ref. \cite{Yordanov}, shown in Fig. \ref{fig_yordanov}, for simulating a four-elements one-qubit POVM.}
\label{fig_yordanov_n4}
\end{figure*}

As a general one-qubit unitary transformation can be recast in terms of four angles as \cite{Nielsen} $\begin{bmatrix} e^{i(\alpha-\beta/2-\delta/2)}\cos\frac{\gamma}{2} & -e^{i(\alpha-\beta/2+\delta/2)}\sin\frac{\gamma}{2} \\ e^{i(\alpha+\beta/2-\delta/2)}\sin\frac{\gamma}{2} & e^{i(\alpha+\beta/2+\delta/2)}\cos\frac{\gamma}{2} \end{bmatrix}$, we see that, given the matrices for the POVM elements in the left hand side of Eq. (\ref{eq:Mjs}), the implementation of the algorithm of Ref. \cite{Yordanov} involves the solution of an intricate system of nonlinear equations for the angles appearing on the right hand side of Eq. (\ref{eq:Mjs}). Perhaps this complication is related to the wrong examples presented in Ref. \cite{Yordanov}, for which the measurement operators do not satisfy the completeness restriction.
It is worthwhile mentioning also that the protocol of Ref. \cite{Yordanov}, if implemented exactly as in the quantum circuit of Fig. \ref{fig_yordanov}, requires $\mathcal{O}(n)$ auxiliary qubits. For diminishing this number to $\mathcal{O}(\log_{2}n)$, one has to make some modifications/additions to this quantum circuit, as exemplified in Fig. \ref{fig_yordanov_n4} for $n=4$.

Quantum state preparation (QSP) algorithms are used as subroutines for performing many tasks \cite{kitaev,shende,plesch,arrazola,araujo,he,zhang,veras}, as for example for implementing the general quantum Fourier transform \cite{kitaev}. Motivated by the issues just discussed about the POVM simulation algorithm of Ref. \cite{Yordanov}, here we present a simple protocol that implements POVMs on any discrete quantum system $A$ associated with a Hilbert space $\mathcal{H}_{A}$ through QSP algorithms. 
For the dimension $d_A$ of the system $A$ on which the POVM is to be implemented and any number of elements of the POVM, if the measurement operators $M_j$ are known, in principle it is possible to calculate the right hand side of Eq. (\ref{eq:iso}): \begin{equation}
\ket{\Psi}_{AB} = \sum_{j}\big(M_{j}|k\rangle_{A}\big)\otimes|j-1\rangle_{B}.
\label{eq:psiAB}
\end{equation}
Once obtained this vector, we can use QSP algorithms to prepare it. Afterwards, a projective measurement on the basis $\{\ket{j}_B\}$ is done. Running several times this procedure, we can extract the probabilities.
%and obtain the post-measurement state, after post-selection of $\ket{j}_B$. 
It is worthwhile mentioning that, since the projective measurements are done on the system $B$, by applying post-selection we can use quantum state tomography to obtain the post-measurement state of the system $A$. 
Therefore, our protocol can be summarized as follows:
\begin{enumerate}
     \item Given $\{M_j\}$, obtain $\ket{\Psi}_{AB}$ of Eq. (\ref{eq:psiAB}).
     \item Implement $\ket{\Psi}_{AB}$ using algorithms for quantum state preparation.
     \item Make projective measurements on system $B$ and extract the measurement statistics.
     \item If needed, implement quantum state tomography to obtain the system $A$ post-measurement state, with post-selection of the measurement results on system $B$.
 \end{enumerate}
This protocol works for any dimension of the system $A$. The minimum dimension of the auxiliary system $B$ is equal to the number of POVM elements, independently of the dimension of the system $A$, i.e., our protocol uses $\mathcal{O}(\log_{2}n)$ auxiliary qubits. Of course, for implementing our protocol on quantum computers based on qubits, it is necessary to choose how to codify the qudit states in terms of qubit states. 
 
\begin{figure}[b!]
\centering
\includegraphics[width=0.48\textwidth]{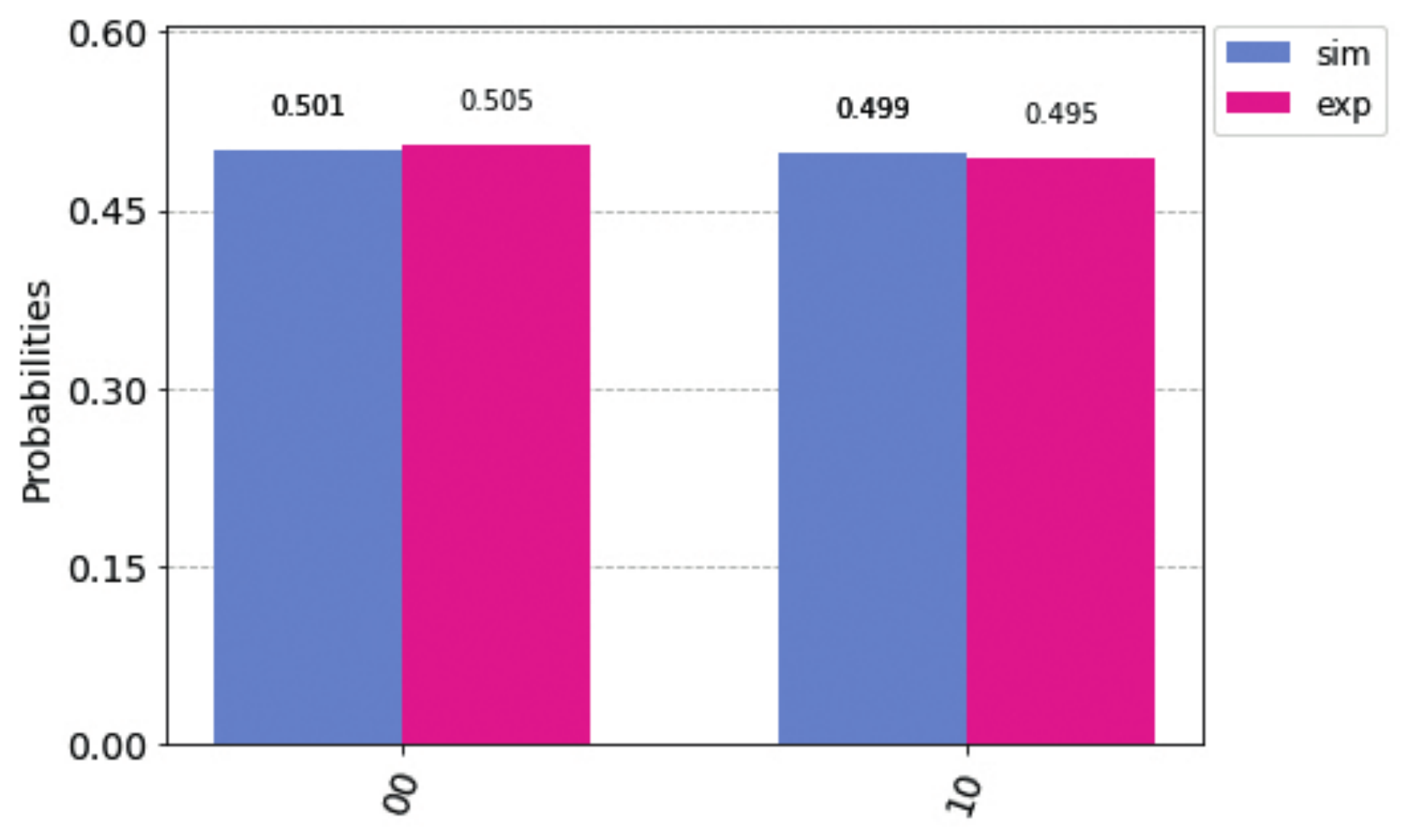}
\caption{Simulation (sim) and demonstration (exp) statistics for the one-qubit two-element POVM of Eq. (\ref{eq:ex1}) implemented using our protocol, for the qubit prepared in the state $|0\rangle_{A}$.}
\label{Ex1}
\end{figure}

In the sequence we present examples of applications of our protocol. Let us start by considering a one-qubit POVM with two elements:
\begin{align}
M_{1}=\frac{1}{2\sqrt{2}}\left(|0\rangle\langle0|+\sqrt{3}|1\rangle\langle0|+2|1\rangle\langle1|\right),  \\
M_{2}=\frac{1}{2\sqrt{2}}\left(|0\rangle\langle0|-\sqrt{3}|1\rangle\langle0|+2|1\rangle\langle1|\right).
\label{eq:ex1}
\end{align}
%We can implement our algorithm from Eq. (\ref{eq:iso})  by choosing an orthonormal basis to encode the auxiliary qubit associated with system $B$ and establishing the state of the target qubit $|\psi_{A}\rangle$. In this case, we consider the computational base ${|j\rangle_{B}}$ and the state $|\psi_{A}\rangle=|0\rangle_{A}$. 
We set the pre-measurement state of system $A$ to $|0\rangle_{A}$. So the POVM probabilities are given by 
\begin{align}
Pr\left(M_{1}|0\right)=\langle0|M_{1}^{\dagger}M_{1}|0\rangle=1/2,\\
Pr\left(M_{2}|0\right)=\langle0|M_{2}^{\dagger}M_{2}|0\rangle=1/2.
\end{align}
In this case we use a qubit as the auxiliary system $B$. For implementing our protocol, we have to prepare the state
\begin{align}
|\Psi_{AB}\rangle=2^{-3/2}\left(|00\rangle+\sqrt{3}|10\rangle+|01\rangle-\sqrt{3}|11\rangle\right).
\end{align}
Here we use the algorithm of Ref. \cite{shende} for QSP. This algorithm was already implemented in Qiskit \cite{qiskit}. 
After state preparation, a projective measurement is performed in the basis $\{|0\rangle_{B},|1\rangle_{B}\}$.
%auxiliary system to obtain the probabilities of each POVM element, given that the target state is $|0\rangle_{A}$. therefore we generate the following histogram
For performing the demonstrations, we used the 
IBMQ \cite{ibmq} quantum chip \textit{ibmq\_belem}.  
The simulation and demonstration results for this first example are shown in Fig. \ref{Ex1}. Some relevant calibration parameters of the quantum chips used for the demonstrations reported in this article are presented in the Appendix.

\begin{figure}[t!]
\centering
\includegraphics[width=0.48\textwidth]{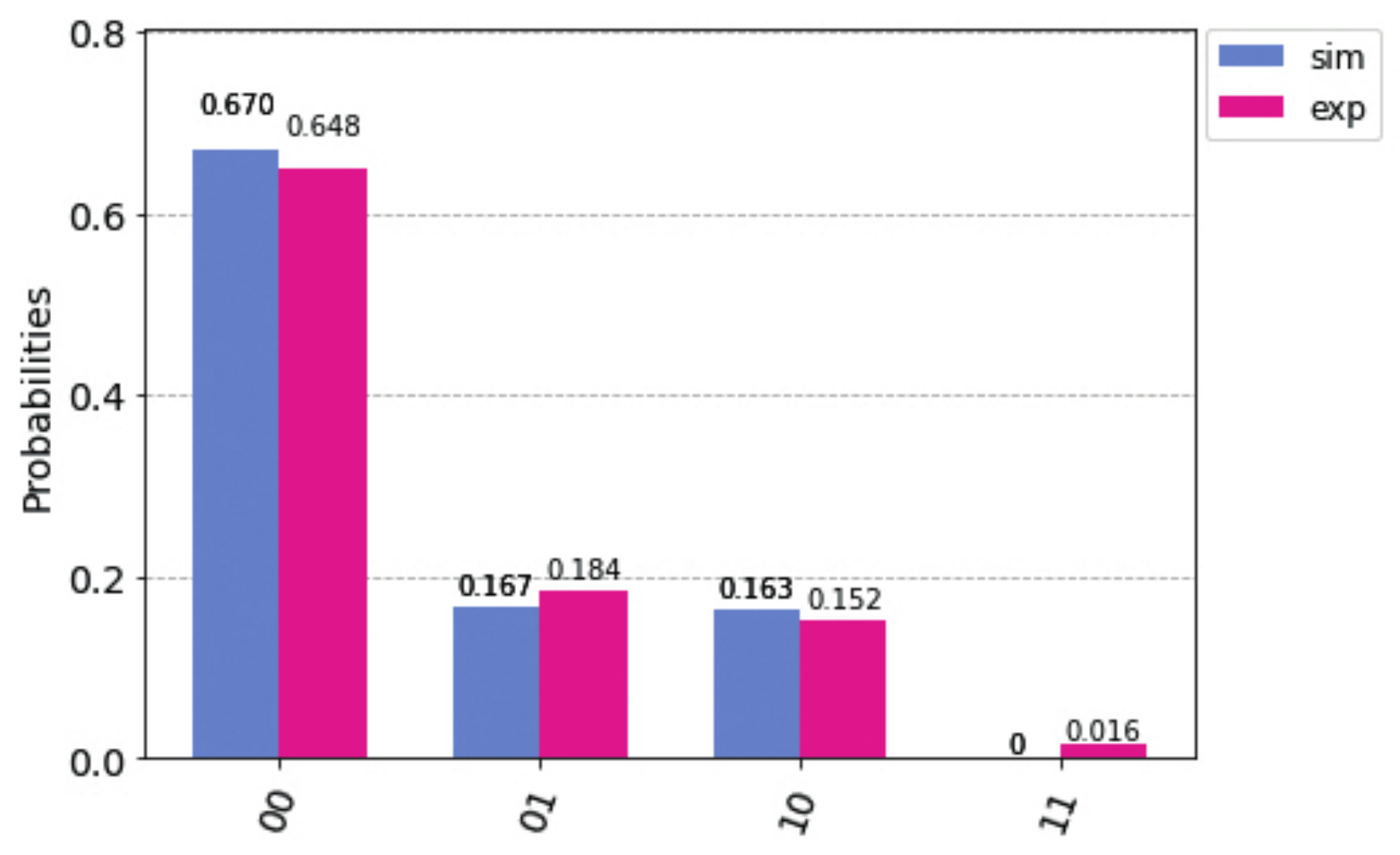}
\caption{Simulation (sim) and demonstration (exp) statistics for the one-qubit three-element POVM of Eq. (\ref{eq:ex2}) implemented using our protocol, for the qubit prepared in the state $|0\rangle_{A}$.}
\label{fig:ex2}
\end{figure}

As a second example, let us consider a one-qubit three-element POVM with measurement elements associated with the sequence of states in the $xz$ plane of the Bloch sphere separated by $2\pi/3$ radians:
\begin{align}
& M_{1}=\sqrt{\frac{2}{3}}|0\rangle\langle0|,  \\
& M_{2}=\sqrt{\frac{2}{3}}|\psi(2\pi/3,0)\rangle\langle\psi(2\pi/3,0)|,  \\
& M_{3}=\sqrt{\frac{2}{3}}|\psi(4\pi/3,0)\rangle\langle\psi(4\pi/3,0)|,
\label{eq:ex2}
\end{align}
with $|\psi\left(\theta,\phi\right)\rangle=\cos\left(\frac{\theta}{2}\right)|0\rangle+\sin\left(\frac{\theta}{2}\right)e^{i\phi}|1\rangle.$ The implementation of this POVM follows the same recipe as for the previous example. In this case the global state to be prepared is
\begin{align}
|\Psi_{Abc}\rangle=& \frac{1}{4}\sqrt{\frac{2}{3}}\left(|0\rangle_{A}\otimes|01\rangle_{bc}+|0\rangle_{A}\otimes|10\rangle_{bc}\right) \\ \nonumber
&+ \sqrt{\frac{2}{3}}|0\rangle_{A}\otimes|00\rangle_{bc} \\ \nonumber
&+\frac{\sqrt{2}}{4}\left(|1\rangle_{A}\otimes|01\rangle_{bc}-|1\rangle_{A}\otimes|10\rangle_{bc}\right),
\end{align}
where we used the qubits $b$ and $c$ to encode the states of the qutrit $B$.
The probabilities, given that the qubit is prepared in the $|0\rangle_{A}$, are 
\begin{align}
Pr\left(M_{1}|0\right)=\langle0|M_{1}^{\dagger}M_{1}|0\rangle=2/3,\\
Pr\left(M_{2}|0\right)=\langle0|M_{2}^{\dagger}M_{2}|0\rangle=1/6.\\
Pr\left(M_{3}|0\right)=\langle0|M_{3}^{\dagger}M_{3}|0\rangle=1/6.
\end{align}
After state preparation, a projective measurement is performed on the basis $\{|00\rangle_{bc},|01\rangle_{bc},|10\rangle_{bc},|11\rangle_{bc}\}$ of the auxiliary system. The obtained probabilities are shown in Fig. \ref{fig:ex2}.

\begin{figure}[t!]
\centering
\includegraphics[width=0.49\textwidth]{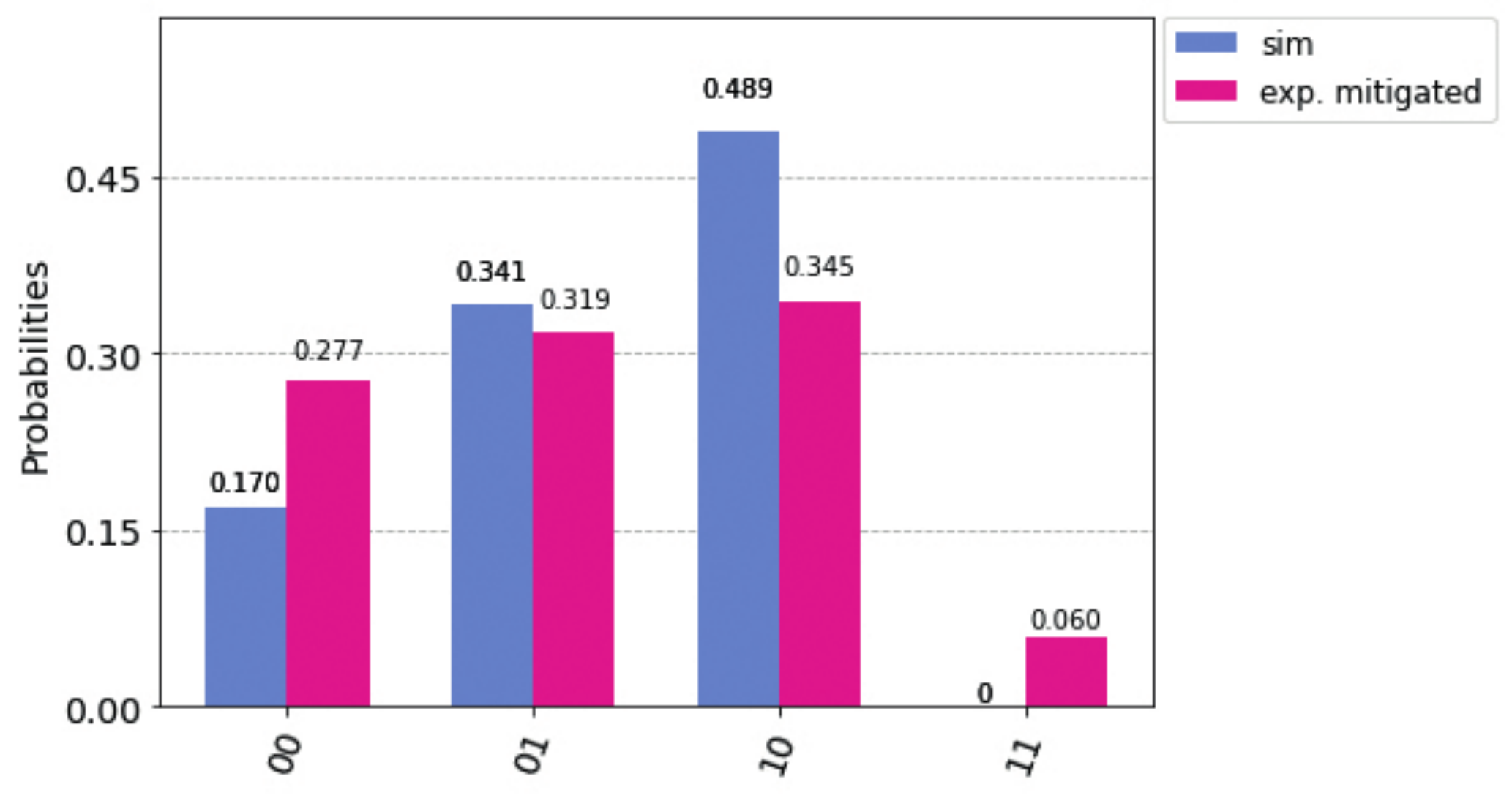}
\caption{Simulation (sim) and demonstration (exp. mitigated) statistics with mitigated errors from \textit{ibmq\_belem} quantum system \cite{ibmq} for the one-qutrit three-element POVM of Eq. (\ref{eq:ex3})  implemented using our protocol, for the qutrit prepared in the state $|\psi_{0}\rangle_{A}$.}
\label{fig:ex3}
\end{figure}

\begin{figure}[t!]
\centering
\includegraphics[width=0.49\textwidth]{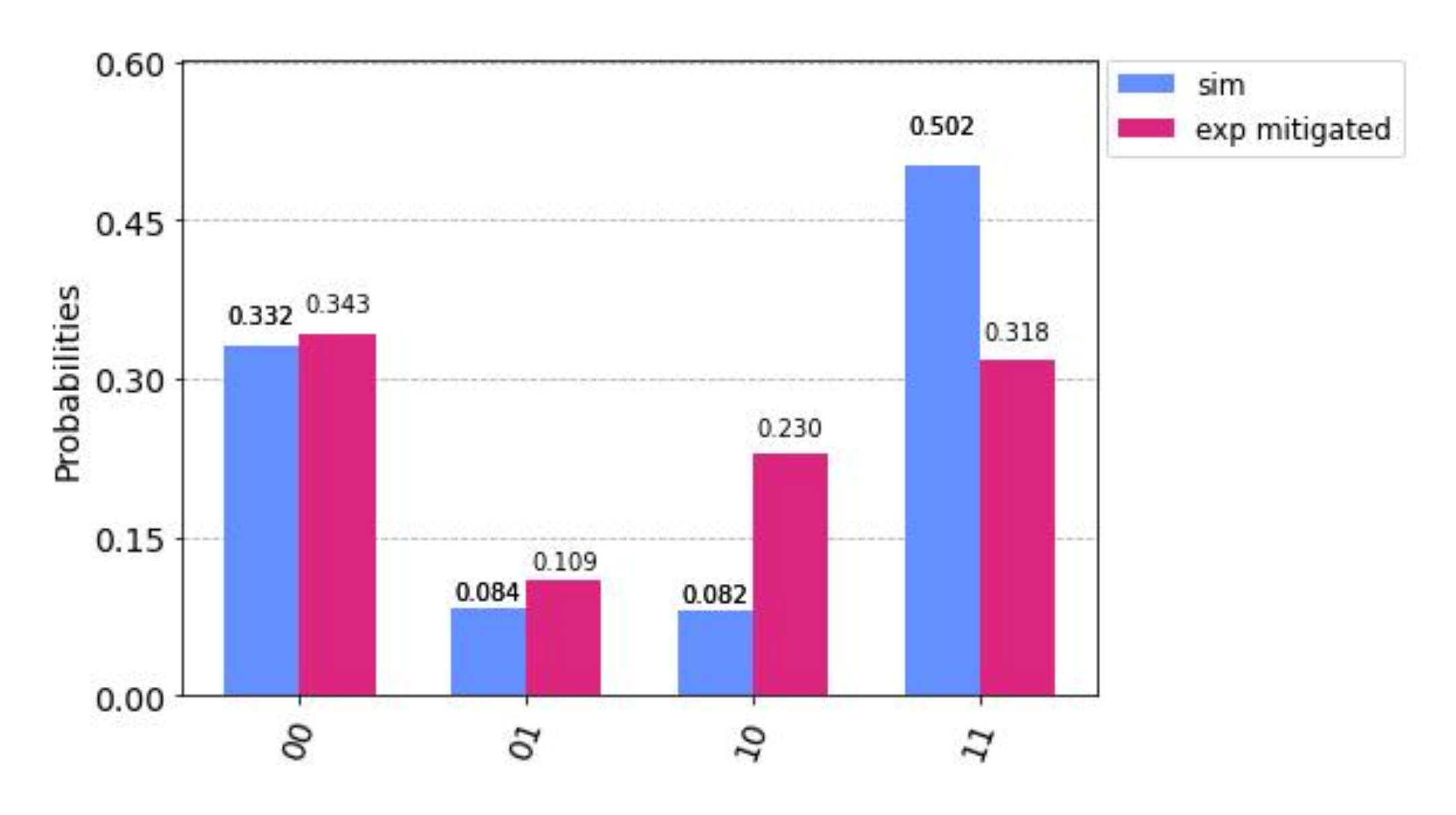}
\caption{Simulation (sim) and demonstration (exp mitigated) statistics with mitigated errors from \textit{ibmq\_belem} quantum system \cite{ibmq} for the two-qubit four-element POVM of Eq. (\ref{eq:ex4})  implemented using our protocol, for the two qubits prepared in the state $|00\rangle_{AB}$.}
\label{fig:ex4}
\end{figure}

As a third example, let us consider a one-qutrit three-element POVM given by
\begin{align}
\label{eq:ex3}
M_{1} =& \frac{1}{2}(|0\rangle+|2\rangle)(\langle0|+\langle2|), \\ 
M_{2} =& \frac{1}{2}(|0\rangle-|2\rangle)(\langle0|-\langle2|), \\ 
M_{3} =& |1\rangle\langle1|.
\end{align}
We set the pre-measurement state of system $A$ to $|\psi_{0}\rangle_{A}=\frac{1}{\sqrt{3}}(|0\rangle+e^{2i\pi/3}|1\rangle+e^{4i\pi/3}|2\rangle)$. In this case the global state to be prepared is
\begin{align}
|\Psi\rangle_{AB} = &	\alpha\left(|0\rangle+|2\rangle\right)_{A}\otimes|0\rangle_{B}+\beta\left(|0\rangle-|2\rangle\right)_{A}\otimes|1\rangle_{B} \nonumber \\ 
	&-\gamma|1\rangle_{A}\otimes|2\rangle_{B},
\end{align} with $\alpha = \frac{\sqrt{3}-3i}{12},$ $\beta = \frac{\sqrt{3}+i}{4}$ and $\gamma = \frac{\sqrt{3}-3i}{6}.$ Using qubits $a$ and $b$ to encode the states of target qutrit $A$ and qubits $c$ and $d$ to encode the states of qutrit $B$, the state vector above can be represented by
\begin{align}
|\Psi\rangle_{abcd} = &	\alpha\left(|00\rangle+|10\rangle\right)_{ab}\otimes|00\rangle_{cd} -\gamma|01\rangle_{ab}\otimes|10\rangle_{cd} \nonumber \\ 
&+\beta\left(|00\rangle-|10\rangle\right)_{ab}\otimes|01\rangle_{cd}.
\end{align}
In this case, given the pre-measurement state $|\psi_{0}\rangle_{A}$ above, the probabilities are 
\begin{align}
Pr\left(M_{1}|\psi_{0}\right)=& \langle\psi_{0}|M_{1}^{\dagger}M_{1}|\psi_{0}\rangle=1/6, \\ 
Pr\left(M_{2}|\psi_{0}\right)=& \langle\psi_{0}|M_{2}^{\dagger}M_{2}|\psi_{0}\rangle=1/2, \\ 
Pr\left(M_{3}|\psi_{0}\right)=& \langle\psi_{0}|M_{3}^{\dagger}M_{3}|\psi_{0}\rangle=1/3.  
\end{align}
As in the previous example, after state preparation, a projective measurement is performed on the basis $\{|00\rangle_{cd},|01\rangle_{cd},|10\rangle_{cd},|11\rangle_{cd}\}$ of the auxiliary system, and the obtained probabilities are shown in Fig. \ref{fig:ex3}.

As a last example, let us consider a two-qubit four element POVM, with measurement operators given as follows
\begin{align}
\label{eq:ex4}
M_{1}	=& \sqrt{\frac{2}{3}}|\Phi_{+}\rangle\langle\Phi_{+}|,  \\ 
M_{2}	=&\sqrt{\frac{2}{3}}|\Psi\left(2\pi/3\right)\rangle\langle\Psi\left(2\pi/3\right)|, \\ 
M_{3}	=&\sqrt{\frac{2}{3}}|\Psi(4\pi/3)\rangle\langle\Psi(4\pi/3)|, \\  
M_{4}	=&|\Phi_{-}\rangle\langle\Phi_{-}|+|\Psi_{-}\rangle\langle\Psi_{-}|,
\end{align} with $|\Psi\left(\theta\right)\rangle=\cos\left(\frac{\theta}{2}\right)|\Phi_{+}\rangle+\sin\left(\frac{\theta}{2}\right)|\Psi_{+}\rangle$.
%i.e, the first three measurement elements are associated with the sequence of Bell basis subspace states $\left\{ |\Phi_{+}\rangle,|\Psi_{+}\rangle\right\}$, separated by $2\pi/3$ radians, while the last element involves the complementary subspace. 
The pre-measurement state of the system $AB$ is given by $|\psi_{0}\rangle_{AB}=|00\rangle$, so the global state to be prepared is
\begin{align}
|\Phi_{abcd}\rangle=& \sqrt{\frac{1}{6}}\left(|00\rangle_{ab}\otimes|00\rangle_{cd}+|11\rangle_{ab}\otimes|00\rangle_{cd}\right) \nonumber \\ &+ \frac{1}{4\sqrt{6}}\left(|00\rangle_{ab}\otimes|01\rangle_{cd}+|11\rangle_{ab}\otimes|01\rangle_{cd}\right) \nonumber \\
	&+ \frac{1}{4\sqrt{2}}\left(|01\rangle_{ab}\otimes|01\rangle_{cd}+|10\rangle_{ab}\otimes|01\rangle_{cd}\right) \nonumber \\ &+ \frac{1}{4\sqrt{6}}\left(|00\rangle_{ab}\otimes|10\rangle_{cd}+|11\rangle_{ab}\otimes|10\rangle_{cd}\right) \nonumber \\
	&-\frac{1}{4\sqrt{2}}\left(|01\rangle_{ab}\otimes|10\rangle_{cd}+|10\rangle_{ab}\otimes|10\rangle_{cd}\right) \nonumber \\ &+\frac{1}{2}\left(|00\rangle_{ab}\otimes|11\rangle_{cd}-|11\rangle_{ab}\otimes|11\rangle_{cd}\right),
\end{align} where we use the qubits $b$ and 
$c$ to encode the states of
the auxiliary ququart system. For the pre-measurement state $|\psi_{0}\rangle_{AB}$ above, we have the following probabilities
\begin{align}
Pr\left(M_{1}|00\right)=&	\langle00|M_{1}^{\dagger}M_{1}|00\rangle=1/3,\\
Pr\left(M_{2}|00\right)=&	\langle00|M_{2}^{\dagger}M_{2}|00\rangle=1/12,\\
Pr\left(M_{3}|00\right)=&	\langle00|M_{3}^{\dagger}M_{3}|00\rangle=1/12,\\
Pr\left(M_{4}|00\right)=&	\langle00|M_{4}^{\dagger}M_{4}|00\rangle=1/2.
\end{align} 
As in the previous examples, after state preparation a projective measurement is performed on the computational basis of the auxiliary system $\{|00\rangle_{cd},|01\rangle_{cd},|10\rangle_{cd},|11\rangle_{cd}\}$, allowing the extraction of the probabilities presented in Fig.  \ref{fig:ex4}.

\begin{figure}[t!]
\centering
\includegraphics[width=0.47\textwidth]{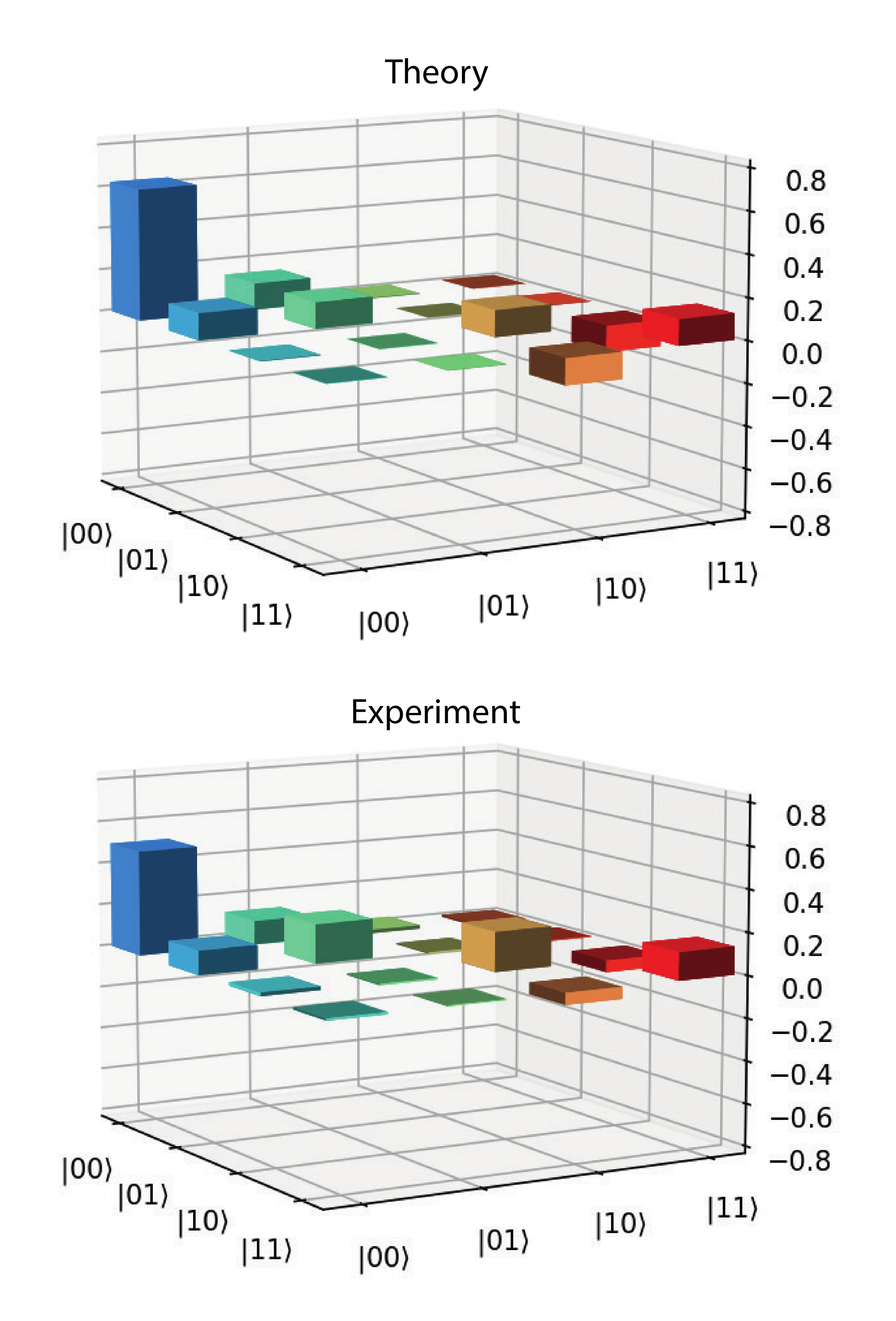}
\caption{Theoretical (a) and demonstration (b) results for the state tomography of the state (quantum instrument) in Eq. (\ref{eq:qi_ex}) simulated using the protocol we introduced in this article.}
\label{fig:tomo_sim_exp}
\end{figure}

Now, let us show how our protocol can be used for implementing \textit{quantum instruments} (QI), that are quantum operations having as input a quantum state and as output a quantum state and a classical variable \cite{Wilde}:
\begin{equation}
\Gamma(|\psi_0\rangle_{A}) = \sum_{j}\varepsilon_{j}(|\psi_0\rangle_{A})\otimes|j\rangle_{J}\langle j|,
\label{eq:qi}
\end{equation}
in which $\{|j\rangle_J\}$ is an orthonormal basis for the system $J$ and 
\begin{equation}
\varepsilon_{j}(|\psi_0\rangle_{A}) = \sum_{k}M_{j,k}|\psi_0\rangle_{A}\langle\psi_0|M_{j,k}^{\dagger}
\end{equation}
is a trace non-increasing quantum operation, i.e., $\Tr(\varepsilon_{j}(|\psi_0\rangle_{A}))\le 1$. Above, $M_{j,k}$ are the elements of a POVM, i.e., $\sum_{j,k}M_{j,k}^\dagger M_{j,k}=\mathbb{I}$. One can verify that the quantum instrument in Eq. (\ref{eq:qi}) can be obtained from the purification
\begin{equation}
|\Psi_{AJE_{j}E_{J}}\rangle = \sum_{j,k}M_{j,k}|\psi_0\rangle_{A}\otimes|j\rangle_{J}\otimes|k\rangle_{E_{j}}\otimes|j\rangle_{E_{J}}.
\end{equation}
That is to say, $\Gamma(|\psi_0\rangle_{A}) = \Tr_{E_j E_J}\big(|\Psi\rangle_{AJE_j E_k}\langle\Psi|\big)$.
So, given the QI, that is to say, given the completely positive maps $\varepsilon_j$ in terms of the set of measurement operators $\{M_{j,k}\}_k$, we can simulate this QI by preparing the state $|\Psi_{AJE_{j}E_{J}}\rangle$ and by taking the partial trace over the auxiliary systems $E_j,E_J$. By measuring the system $J$ in the basis $\{|j\rangle_J\}_j$ and post-selecting the results, we can also reconstruct the action of the operators $\varepsilon_j(|\psi_0\rangle_A)$. 
In what follows, we exemplify the application of this simulation protocol. Let us consider a one-qubit QI defined by the following set of trace non-increasing quantum operations:
\begin{align}
& \varepsilon_0\equiv\Big\{M_{00}= \frac{1}{\sqrt{2}}|0\rangle\langle0|, M_{01}=	\frac{1}{\sqrt{2}}|+\rangle\langle+|\Big\},  \\
& \varepsilon_1 \equiv \Big\{M_{10}=	\frac{1}{\sqrt{2}}|1\rangle\langle1|, 
M_{11}=	\frac{1}{\sqrt{2}}|-\rangle\langle-|\Big\}.
\end{align}
We set the pre-measurement state of system $A$ to $|\psi_{0}\rangle_{A}=|0\rangle_{A}$. The global state to be prepared for the simulation of this QI is
\begin{align}
\sqrt{2}|\Psi\rangle_{AJE_{j}E_{J}}=&	|0000\rangle_{AJE_j E_J}+\frac{1}{2}|0010\rangle_{AJE_j E_J} \nonumber \\
& + \frac{1}{2}|1010\rangle_{AJE_j E_J} + \frac{1}{2}|0111\rangle_{AJE_j E_J} \nonumber \\
& - \frac{1}{2}|1111\rangle_{AJE_j E_J}.
\end{align}
%so we have the probabilities 
%\begin{align}
%Pr(M_{00}|0) & = \langle 0|M_{00}^{\dagger}M_{00}|0\rangle = \langle0|\left(\frac{1}{2}|0\rangle\langle0|\right)|0\rangle = 1/2 \\
%Pr(M_{01}|0) & = \langle 0|M_{01}^{\dagger}M_{01}|0\rangle = \langle0|\left(\frac{1}{2}|+\rangle\langle+|\right)|0\rangle =1/4, \\
%Pr(M_{10}|0) & = \langle 0|M_{10}^{\dagger}M_{10}|0\rangle = \langle0|\left(\frac{1}{2}|1\rangle\langle1|\right)|0\rangle =0, \\
%Pr(M_{11}|0) & = \langle 0|M_{11}^{\dagger}M_{11}|0\rangle = \langle0|\left(\frac{1}{2}|-\rangle\langle-|\right)|0\rangle =1/4. \\ \nonumber
%\end{align}
From this quantum state, we obtain the quantum instrument
\begin{align}
    & 8\Gamma(|0\rangle_A) = 8\Tr_{E_{j}E_{J}}\left\{ |\Psi\rangle_{AJE_j E_J}\langle\Psi|\right\} \label{eq:qi_ex} \\
    &= 	\left(5|0\rangle_A \langle0|+|0\rangle_A \langle1|+|1\rangle_A \langle0|+|1\rangle_A \langle1|\right)|0\rangle_J \langle0| \nonumber \\
	&+\left(|1\rangle_A \langle1|-|0\rangle_A \langle1|-|1\rangle_A \langle0|+|0\rangle_A \langle0|\right)|1\rangle_J \langle1| \nonumber \\
 & = \varepsilon_0(|0\rangle_A)\otimes|0\rangle_B \langle 0| + \varepsilon_1(|0\rangle_A)\otimes|1\rangle_B \langle 1|. \nonumber
\end{align}
This state was reconstructed using quantum state tomography. The theoretical and demonstration results are shown in Fig. \ref{fig:tomo_sim_exp}. For this demonstration we use the IBM quantum chip \textit{ibmq\_belem}  \cite{ibmq}. In this case, the obtained demonstration result also agreed quite well with the theoretical prediction. 

%\begin{figure}%[H]%
%\centering
%\label{fig:first}%
%\includegraphics[height=2.9in]{tomo_qinstrument.png}%}%
%\qquad
%\label{fig:second}%
%\includegraphics[height=2.9in]{tomo_exp.png}%}%
%\caption{Theoretical and experimental results for the state tomography of the state in Eq. (\ref{eq:qi_ex}) simulated using the protocol introduced in this article.}
%\label{fig:tomo_sim_exp}
%\end{figure}

%\begin{figure}[t!]
%\centering
%\includegraphics[width=0.47\textwidth]{tomo_pt_exp.png}
%\caption{experimental tomography  statistics  from \textit{ibmq\_belem} quantum system \cite{ibmq} for the one-qubit four-element POVM of quantum instrument implemented using our protocol, and the  target qubit prepared in the state $|0\rangle_{A}$.}
%\label{fig:tomo_exp}
%\end{figure}

In summary, 
%we identified basic errors on the examples used, 
we pointed out that changes in the algorithm are needed for maintaining the claimed scaling of the number of auxiliary qubits, and we highlighted the practical difficulties of the protocol presented in Ref. \cite{Yordanov}. We circumvented these difficulties through the use of quantum state preparation algorithms. 
This approach avoids numerical and computational issues associated with the solution of systems of nonlinear equations, and easily generalizes the implementation of Ref. \cite{Yordanov} for $2$-level states (one qubit) to $d$-level states, avoiding also the complications of their algorithm regarding the implementation of general unitary transformations on a multiqubit system. 
%We hope that our protocol will contribute to accelerate the research using POVMs.
We exemplified the application of our protocol for one qubit, one qutrit and two qubit POVMs and for simulating quantum instruments. These examples were demonstrated using IBM quantum computers. The simulation results matched the theoretical predictions. The demonstration results are in fairly good agreement with theory, but can be further improved if this protocol is executed in lesser noise quantum devices. 
So, we believe that the simplicity and easy of use of this protocol will foster further research involving POVMs.

%At this point, it is worthwhile to mention that, in the algorithm of Ref. \cite{Yordanov}, for each measurement operator added, one has to add one more auxiliary qubit. This is easy to see from Fig. \ref{fig_yordanov}. On the other hand, in our algorithm the dimension of the auxiliary system is equal to the number of POVM elements. So, for example, for a one-qubit eight-element POVM, their algorithm uses seven auxiliary qubits while our uses only three. If we would like to implement a one-qubit sixteen-elements POVM, their algorithm would require fifteen auxiliary qubits while ours needs four auxiliary qubits. Generally, their algorithm involves $\mathcal{O}(N)$ and ours $\mathcal{O}(\log_{2}N)$ auxiliary qubits, at least for when number of POVM elements $N$ is a power of two.
 
\begin{acknowledgments}
This work was supported by the by the S\~ao Paulo Research Foundation (FAPESP), Grant No.~2022/09496-8, by the National Institute for the Science and Technology of Quantum Information (INCT-IQ), process 465469/2014-0, by the Coordination for the Improvement of Higher Education Personnel (CAPES), process 88882.427913/2019-01, by the National Council for Scientific and Technological Development (CNPq), process 309862/2021-3, and by the Brazilian Space Agency (AEB), process 01350.001732/2020-61 (TED 020/2020).
\end{acknowledgments}

%-------------------------------
\appendix*
\section{Information about the used quantum chip}
\label{sec:ap}

\begin{figure}[ht]
\centering
\includegraphics[width=0.2\textwidth]{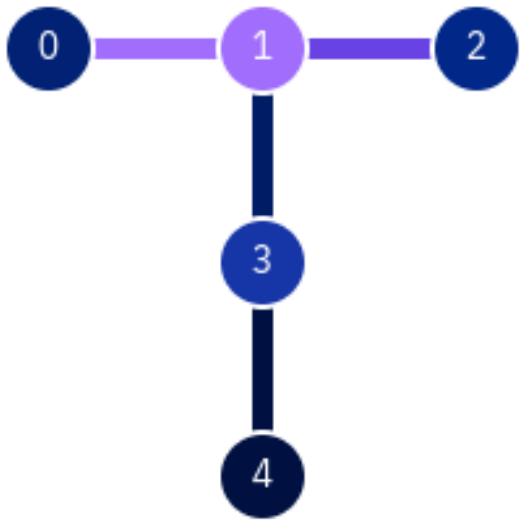}%{ibmq_belem_calibrations_readout_error_cx_map_2023-01-25T18_10_48Z.png}
\caption{Illustration of the connectivity between qubits of the \textit{ibmq\_belem} quantum chip.}
\label{fig:ibmq_belem_connect}
\end{figure}

In this article, we implement our POVM simulation algorithm using IBMQ platform \cite{ibmq}. We use quantum chips through Qiskit, an Open Source Quantum Development Kit for working with quantum computers at the level of pulses, circuits and application modules.
In our demonstrations, we used the \textit{ibmq\_belem} quantum chip with the same calibration parameters presented in Table \ref{tab:table1} for the examples in Figs. \ref{Ex1}, \ref{fig:ex2}, \ref{fig:ex3}, and \ref{fig:ex4}. On the other hand, for the example in Fig. \ref{fig:tomo_sim_exp}, we used the same chip but with calibration parameters as shown in Table \ref{tab:table2}. The \textit{ibmq\_belem} quantum chip connectivity is shown in Fig. \ref{fig:ibmq_belem_connect}.

\begin{table}[ht]
\caption{\label{tab:table1}
Calibration parameters for the \textit{ibmq\_belem}  quantum chip when used for the examples in Figs. \ref{Ex1}, \ref{fig:ex2}, \ref{fig:ex3} and \ref{fig:ex4}.}
\begin{ruledtabular}
\begin{tabular}{cccccccc}
&Q0&Q1&Q2&Q3&Q4\\
\hline
Frequency (GHz)& 5.09 & 5.246 & 5.361 & 5.17 & 5.259  \\
T1 ($\mu$s)& 155.98 & 119.12 & 91.07 & 63.29 & 88.85 \\
T2 ($\mu$s)& 121.74 & 117.28 & 58.25 & 146.89 & 151.6 \\
1-qubit error ($10^{-4}$)& 1.69 & 17.87
& 2.57 & 6.36 & 13.64 \\
Readout error ($10^{-2}$)& 1.45 & 2.90 & 2.65 & 3.78 & 3.26 \\
CNOT error ($10^{-2}$)& $\begin{array}{c}
0-1\\
1.858
\end{array}$ & $\begin{array}{c}
1-3\\
1.293
\end{array}$& $\begin{array}{c}
2-1\\
1.089
\end{array}$& $\begin{array}{c}
3-4\\
2.143
\end{array}$ & $\begin{array}{c}
4-3\\
2.143
\end{array}$\\
&  & $\begin{array}{c}
1-2\\
1.089
\end{array}$&  &$\begin{array}{c}
3-1\\
1.293
\end{array}$  &  \\
&  & $\begin{array}{c}
1-0\\
1.858
\end{array}$& &  &  \\
\end{tabular}
\end{ruledtabular}
\end{table}

\begin{table}[ht]
\caption{\label{tab:table2}
Calibration parameters for the \textit{ibmq\_belem}  quantum chip when used for the example in Fig. \ref{fig:tomo_sim_exp}.}
\begin{ruledtabular}
\begin{tabular}{cccccccc}
&Q0&Q1&Q2&Q3&Q4\\
\hline
Frequency (GHz)& 5.09 & 5.246 & 5.361 & 5.17 & 5.23  \\
T1 ($\mu$s)& 128.46 & 88.52 & 85.16 & 79.29 & 1.55 \\
T2 ($\mu$s)& 127.51 & 99.97 & 65.4 & 134.08 & 92.04 \\
1-qubit error ($10^{-4}$)& 1.651 & 2.937
& 2.912 & 4.044 & 1.849$\times10^{3}$  \\
Readout error ($10^{-2}$)& 1.31 & 1.99 & 2.19 & 2.87 & 14.67 \\
CNOT error ($10^{-2}$)& $\begin{array}{c}
0-1\\
1.345
\end{array}$ & $\begin{array}{c}
1-3\\
1.682
\end{array}$& $\begin{array}{c}
2-1\\
0.698
\end{array}$& $\begin{array}{c}
3-4\\
100.0
\end{array}$ & $\begin{array}{c}
4-3\\
100.0
\end{array}$\\
&  & $\begin{array}{c}
1-2\\
0.698
\end{array}$&  &$\begin{array}{c}
3-1\\
1.682
\end{array}$  &  \\
&  & $\begin{array}{c}
1-0\\
1.345
\end{array}$& &  &  \\
\end{tabular}
\end{ruledtabular}
\end{table}

%------------------------------

\end{document}